\documentclass[journal=nalefd,manuscript=article]{achemso}

\usepackage[version=3]{mhchem} 
\usepackage{xcolor}
\usepackage[normalem]{ulem}

\usepackage{titlesec}
\titleformat{\section}
  {\normalfont\fontsize{12}{13}\bfseries}{\thesection}{1em}{}
\usepackage{caption}
\DeclareCaptionFont{mysize}{\fontsize{9}{9.6}\selectfont}
\usepackage[separate-uncertainty=true,multi-part-units=single,range-phrase=--,range-units=single]{siunitx}

\author{Alan M. Dibos}
    \affiliation[Argonne National Laboratory]
    {Nanoscience and Technology Division, Argonne National Laboratory, Lemont, IL 60439}
    \alsoaffiliation[Argonne National Laboratory]
    {Center for Molecular Engineering, Argonne National Laboratory, Lemont, IL 60439}
    \email{adibos@anl.gov}
\author{Michael T. Solomon}
    \affiliation[University of Chicago]
    {Pritzker School of Molecular Engineering, University of Chicago, Chicago, IL 60637}
    \alsoaffiliation[Argonne National Laboratory]
    {Materials Science Division, Argonne National Laboratory, Lemont, IL 60439}
    \alsoaffiliation[Argonne National Laboratory]
    {Center for Molecular Engineering, Argonne National Laboratory, Lemont, IL 60439}
\author{Sean E. Sullivan}
    \affiliation[Argonne National Laboratory]
    {Materials Science Division, Argonne National Laboratory, Lemont, IL 60439}
    \alsoaffiliation[Argonne National Laboratory]
    {Center for Molecular Engineering, Argonne National Laboratory, Lemont, IL 60439}
\author{Manish K. Singh}
    \affiliation[University of Chicago]
    {Pritzker School of Molecular Engineering, University of Chicago, Chicago, IL 60637}
    \alsoaffiliation[Argonne National Laboratory]
    {Materials Science Division, Argonne National Laboratory, Lemont, IL 60439}
\author{Kathryn E. Sautter}
    \affiliation[Argonne National Laboratory]
    {Materials Science Division, Argonne National Laboratory, Lemont, IL 60439}
    \alsoaffiliation[University of Chicago]
    {Pritzker School of Molecular Engineering, University of Chicago, Chicago, IL 60637}
\author{Connor P. Horn}
    \affiliation[University of Chicago]
    {Pritzker School of Molecular Engineering, University of Chicago, Chicago, IL 60637}
    \alsoaffiliation[Argonne National Laboratory]
    {Materials Science Division, Argonne National Laboratory, Lemont, IL 60439}
\author{Gregory D. Grant}
   \affiliation[University of Chicago]
    {Pritzker School of Molecular Engineering, University of Chicago, Chicago, IL 60637}
    \alsoaffiliation[Argonne National Laboratory]
    {Materials Science Division, Argonne National Laboratory, Lemont, IL 60439}
\author{Yulin Lin}
    \affiliation[Argonne National Laboratory]
    {Nanoscience and Technology Division, Argonne National Laboratory, Lemont, IL 60439}
\author{Jianguo Wen}
    \affiliation[Argonne National Laboratory]
    {Nanoscience and Technology Division, Argonne National Laboratory, Lemont, IL 60439}
\author{F. Joseph Heremans}
    \affiliation[Argonne National Laboratory]
    {Materials Science Division, Argonne National Laboratory, Lemont, IL 60439}
    \alsoaffiliation[Argonne National Laboratory]
    {Center for Molecular Engineering, Argonne National Laboratory, Lemont, IL 60439}
    \alsoaffiliation[University of Chicago]
    {Pritzker School of Molecular Engineering, University of Chicago, Chicago, IL 60637}
\author{Supratik Guha}
    \affiliation[University of Chicago]
    {Pritzker School of Molecular Engineering, University of Chicago, Chicago, IL 60637}
    \alsoaffiliation[Argonne National Laboratory]
    {Materials Science Division, Argonne National Laboratory, Lemont, IL 60439}
\author{David D. Awschalom}
    \affiliation[University of Chicago]
    {Pritzker School of Molecular Engineering, University of Chicago, Chicago, IL 60637}
    \alsoaffiliation[Argonne National Laboratory]
    {Materials Science Division, Argonne National Laboratory, Lemont, IL 60439}
    \alsoaffiliation[Argonne National Laboratory]
    {Center for Molecular Engineering, Argonne National Laboratory, Lemont, IL 60439}
    
\title[An \textsf{achemso} demo]
  {Purcell enhancement of erbium ions in TiO\textsubscript{2} on silicon nanocavities}


\begin{document}







\begin{abstract}
Isolated solid-state atomic defects with telecom optical transitions are ideal quantum photon emitters and spin qubits for applications in long-distance quantum communication networks. Prototypical telecom defects such as erbium suffer from poor photon emission rates, requiring photonic enhancement using resonant optical cavities. Many of the traditional  hosts for erbium ions are not amenable to direct incorporation with existing integrated photonics platforms, limiting scalable fabrication of qubit-based devices. Here we present a scalable approach towards CMOS-compatible telecom qubits by using erbium-doped titanium dioxide thin films grown atop silicon-on-insulator substrates. From this heterostructure, we have fabricated one-dimensional photonic crystal cavities demonstrating quality factors in excess of $5\times10^{4}$ and corresponding Purcell-enhanced optical emission rates of the erbium ensembles in excess of 200. This easily fabricated materials platform represents an important step towards realizing telecom quantum memories in a scalable qubit architecture compatible with mature silicon technologies.
\end{abstract}


\section*{Main Text}
Rare earth ion defects in solid-state hosts are key candidate qubits for applications in quantum computing and communication owing to their inherent spin-photon interface and long coherence times. These properties have enabled critical demonstrations of optical quantum memory protocols~\cite{Lvovsky2009} based on light-matter entanglement\cite{Clausen2011} and entanglement distribution\cite{Lago-Rivera2021}. Quantum technologies geared toward distribution of quantum information over long distances require qubits that interface with telecom photons to avoid huge propagation losses over optical fiber networks. Because the trivalent ion of erbium (Er$^{3+}$) has a $^{4}\text{I}_{15/2}$~$\rightarrow$~$^{4}\text{I}_{13/2}$ transition in the telecom C-band, there has been interest in using erbium ions as optically-addressable quantum memories using persistent spectral hole burning techniques \cite{Saglamyurek2015,Miyazono2016,Craiciu2019} and spin-based \cite{Rancic2018} quantum memories in quantum communication, including at the level of single ions \cite{Dibos2018,Zhong2019,Raha2020}. However, significant engineering steps are needed to enhance the low photon emission rates from individual rare earth ions, which tend to have long radiative lifetimes.

One direct way to reduce the radiative lifetime is to use an optical cavity to enhance the emission through the Purcell effect~\cite{Martini1987}. The majority of Purcell enhancement studies on rare earth ions have involved devices derived from rare earth-doped bulk crystals, namely heterogeneous integration with deposited amorphous Si resonators~\cite{Miyazono2016} or bonded Si photonics \cite{Dibos2018,Chen2021}, focused ion beam milling of bulk crystals \cite{Zhong2017, Craiciu2019, Craiciu2021,Kindem2020}, or incorporating relatively small erbium-doped samples into tunable distributed Bragg reflector-based fiber cavities \cite{Merkel2020, Casabone2021}. These approaches are all appealing because they leverage the generally good performance of well-studied host materials for Er$^{3+}$ (Y\textsubscript{2}SiO\textsubscript{5} \cite{Miyazono2016,Dibos2018, Craiciu2019, Craiciu2021}, YVO\textsubscript{4} \cite{Xie2021}, CaWO\textsubscript{4} \cite{Bertaina2007, LeDantec2021}). Furthermore, these prior approaches have led to important cutting-edge demonstrations towards single rare-earth ion quantum memories, including single-shot spin state readout \cite{Raha2020,Chen2020,Kindem2020}. However, such approaches may not allow for the on-chip scalability needed for wide deployment of quantum memories in large-scale quantum networks. For instance, estimates for fault-tolerant quantum repeater protocols show that hundreds or thousands of communication qubits are necessary to perform the requisite (operational and loss) error correction for quantum communication over long distances ($\ge$1000~km)~\cite{Muralidharan2014}. In addition, the implementation of a fully CMOS-compatible process flow can enable easy integration with on-chip filters, phase shifters, and beamsplitters for on-chip photon routing \cite{Moody2022}. In the vein of technological scalability with Si, there has also been considerable interest on low concentration implantation of erbium ions into Si nanostructures \cite{Weiss2021}. While this approach might have the greatest technological upside for directly unifying Si photonics with erbium-based quantum memories, early work has shown that the size mismatch between Er$^{3+}$ and the crystalline lattice leads to myriad substitutional and interstitial sites with closely spaced spectral emission\cite{Berkman2021,Hu2022}, and further experiments are needed to resolve the the best defects for various applications.

In addition to ease of Si photonic integration---assessed by the ability to grow on Si and perform straightforward fabrication---there are a variety of additional considerations for a candidate material to serve as a high-quality host for Er$^{3+}$ such as a minimal or controlled level of background nuclear spins, a wide bandgap, the number of substitutional sites available, and site symmetries~\cite{Ferrenti2020,Stevenson2021, Kanai2021}. To that end, previous measurements have suggested that rutile TiO\textsubscript{2} would be a good host for erbium because of narrow optical and spin linewidths in implanted bulk crystals owing to the nonpolar symmetry of the substitutional erbium site and the low natural abundance of nuclear spins, respectively \cite{Phenicie2019}. Furthermore, TiO\textsubscript{2} thin films can be grown on Si using a variety of deposition techniques \cite{Zhu2015,Singh2022}, are CMOS-compatible, and are amenable to standard fluorine- and chlorine-based dry etch chemistries for top-down device fabrication \cite{Norasetthekul2001}. Recent work has shown Er-doped TiO\textsubscript{2} thin films grown on Si substrates via molecular beam deposition\cite{Singh2022}. Some salient features from that recent work are that the TiO\textsubscript{2} thin films were polycrystalline and the dominant TiO\textsubscript{2} phase (rutile or anatase) could be tuned by varying the substrate temperature during growth. Furthermore, the inhomogeneous linewidths for the best buffered devices were as low as 5~GHz, which was substantially narrower than some of the epitaxial thin film control samples grown on better lattice-matched substrates (SrTiO\textsubscript{3} or sapphire). Molecular beam deposition is particularly appealing for these types of heterostructures because of the inherent control of the Er$^{3+}$ doping profile and doping density, in order to spectrally resolve single ions. Furthermore, unintentional Er$^{3+}$ doping from the TiO\textsubscript{2} itself is anticipated to be lower than for the more common yttrium-containing oxides due to relatively high concentrations of trace lanthanides in yttrium precursors.

\begin{figure}
    \includegraphics[width=1.0\textwidth]{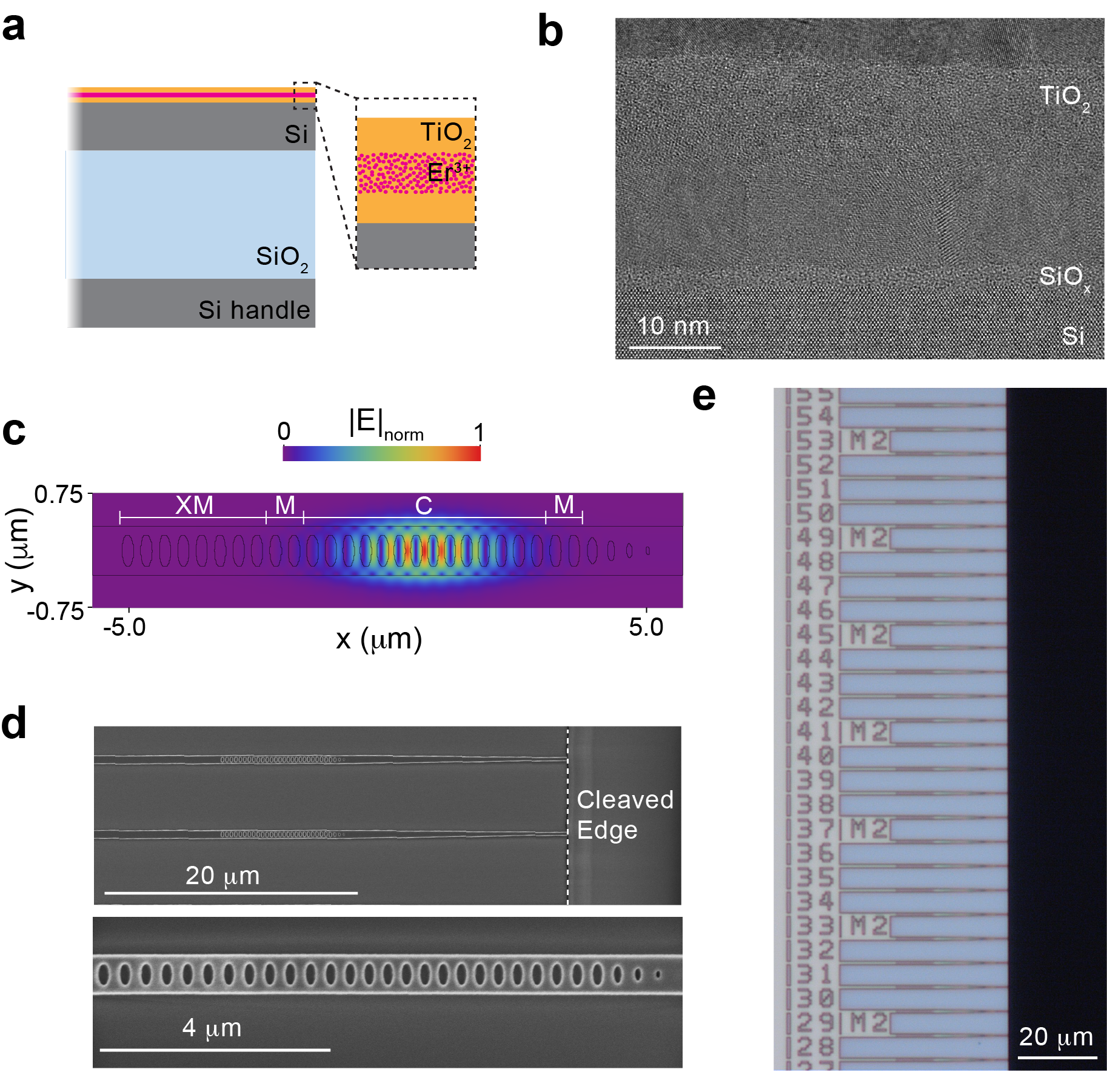}
    \caption{\textbf{Er-doped TiO$_2$ on SOI device platform.} (a)~Illustration of the Er-doped TiO\textsubscript{2} heterostructure. The Er$^{3+}$ layer is sandwiched between nominally undoped layers. The thickness of the SOI layers is not drawn to scale. (b)~Cross-sectional TEM image of polycrystalline TiO\textsubscript{2} film atop the Si layer. A modest SiO\textsubscript{x} layer of approximately 2 nm thickness develops at the interface during growth. (c)~FDTD simulation of the normalized electric field confinement for a dipole oriented normal to the waveguide and the plane of the TiO$_2$ film, located 10~nm above the surface of the Si. Strong electric field confinement in the photonic crystal 14-hole cavity defect (C) is generated by a parabolic taper of the lattice constant of the elliptically shaped holes. There are two mirror holes (M) on either side of the cavity region, and extra mirror holes (XM) are included on the left hand side of the device because all measurements are performed in a one-sided reflection configuration. (d)~Top:~SEM image of an entire fabricated device showing the tapered waveguide extending from the cleaved edge of the SOI chip. Bottom: An expanded view of the parabolic taper in the lattice constant to generate the cavity defect. (e) An optical image showing an extended view of nearly identical devices, including those in (d).}
    \label{fig1}
\end{figure}

In this Letter, we show early progress on top-down nanofabrication of photonic crystal cavities comprising thin layers of Er-doped TiO\textsubscript{2} grown directly atop Si films. The thin films explored are similar to those discussed previously \cite{Singh2022}, but consist of an Er-doped TiO\textsubscript{2} heterostructure grown on commercial silicon-on-insulator (SOI) substrates typically used by the Si photonics community: Si device layer thickness of 220 nm atop a \SI{2}{\um} buried oxide (see Methods). Within each TiO\textsubscript{2} heterostructure there are three TiO\textsubscript{2} layers of equal thickness: an erbium-doped layer sandwiched between undoped top and bottom buffer layers, and each layer is \SI{\sim 7.5}{\nm}, giving a total film thickness of \SI{\sim 22}{\nm} (Fig.~\ref{fig1}a). The estimated Er$^{3+}$ density within the doped layer is 35 ppm (see Methods). The TiO\textsubscript{2} layers for this sample are grown at a substrate temperature of 520~\textdegree C. As shown in Fig.~\ref{fig1}b, transmission electron microscopy (TEM) cross-sectional imaging reveals that the TiO\textsubscript{2} layer is polycrystalline, and there is a thin oxide layer ($\sim$2~nm) at the Si/TiO\textsubscript{2} interface, as was seen previously for lower temperature growth. However, unlike the previous demonstration, the polycrystalline film has both large and small grains visible, and the grains nearest the Si interface are bigger ($13\pm3$~nm) than those near the top of the TiO\textsubscript{2} layer ($\sim$1~nm). The cause of this gradient in grain size and effect on eventual Er properties is currently still under investigation. The roughness of the top surface of the TiO\textsubscript{2} is approximately 1 nm according to the cross-sectional TEM. Using 3D FDTD simulations, we have designed one-dimensional photonic crystal cavities for our Si/TiO\textsubscript{2} heterostructure. Figure~\ref{fig1}c shows the computed fundamental (dielectric) mode with a predominant TE-like polarization in the plane of the TiO\textsubscript{2} film. The photonic crystal devices consist of a waveguide with identical, elliptically shaped holes and a parabolic reduction of the lattice constant to generate a defect in the photonic bandgap~\cite{Dibos2018} (Fig.~\ref{fig1}d, bottom). The photonic crystal cavities are patterned via conventional electron-beam lithography and dry etching through both the TiO\textsubscript{2} and Si device layers (see Methods and Supporting Information). After etching and hardmask removal, the waveguide is cleaved along alignment marks under a microscope, beyond the inverse-taper (Fig.~\ref{fig1}d, top). We employ an end-fire measurement configuration, where a lensed optical fiber is used to couple light in/out of the device. As a demonstration of the inherent scalability of this scheme, we fabricate hundreds of photonic crystal devices on each chip, a representative cluster of which is shown in Fig.~\ref{fig1}e.

\begin{figure}
    \includegraphics[width=3.3in]{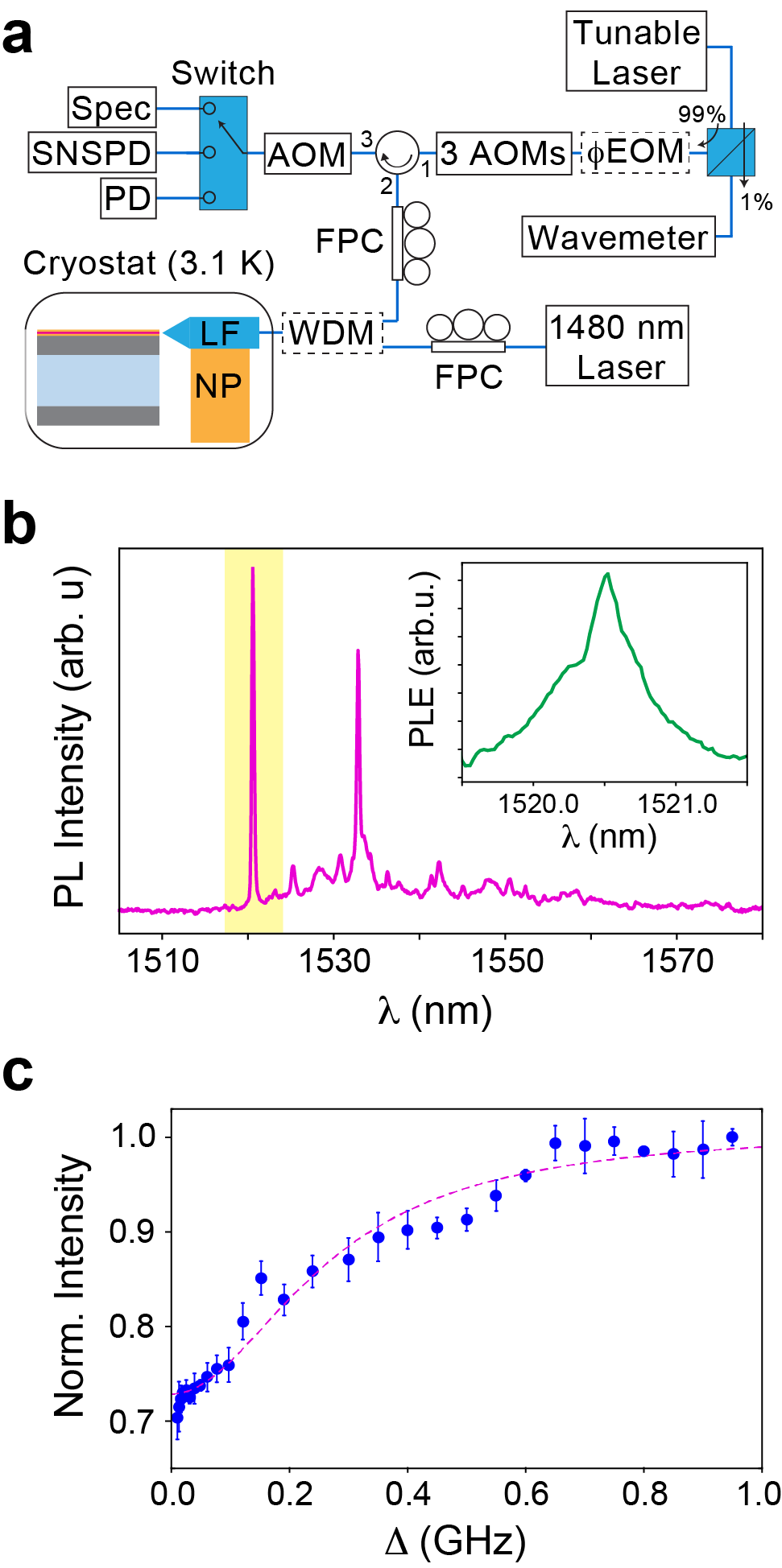}
    \caption{\textbf{Optical characterization of Er$^{3+}$:TiO$_2$ on Si waveguides.} (a)~Schematic of the experimental configuration. The Er$^{3+}$:TiO$_2$ waveguide device is situated in a cryostat ($\text{T} = \SI{3.1}{\K}$). A tunable laser in combination with three fiber-coupled acousto-optic modulators (AOMs) enable the production of short pulses of light that are directed to the sample through a lensed fiber (LF) mounted on a nanopositioner (NP). The return light can be routed to either a photodiode (PD), IR spectrometer (Spec), or superconducting nanowire single-photon detector (SNSPD). A fiber polarization controller (FPC) is used to rotate the polarization to match that of the cavity. We can insert optional components (dashed boxes) such as an electro-optic phase modulator ($\phi$EOM) to generate sidebands for transient spectral hole burning and a fiber wavelength division multiplexer (WDM) for off-resonant excitation. Additional details are given in the Methods section. (b)~PL spectrum of a waveguide-only (no cavity) device pumped with 1480 nm laser light and detected via spectrometer. The tallest peak (highlighted in yellow) is centered at 1520.56~nm and originates from Er$^{3+}$ in the rutile phase of polyphase TiO\textsubscript{2}. Inset: A resonant laser scan showing a similar inhomogeneous linewidth of \SI{0.4}{\nm} (\SI{65}{\GHz}) for the rutile peak. (c)~Measurement of the spectral diffusion linewidth for the rutile transition at 1520.56 nm. The inverted Lorentzian fit (dashed magenta line) yields a spectral diffusion linewidth of \SI{267(17)}{\MHz}.}
    \label{fig2}
\end{figure}

A schematic of our experimental setup used to measure these devices is shown in Fig.~\ref{fig2}a. We perform our measurements in a closed-cycle cryostat with a base temperature of $\text{T} = \SI{3.1}{\K}$. The sample is fixed to the cold finger and the lensed fiber is mounted on a 3-axis nanopositioner to enable addressing of different devices. Using a tunable telecom C-band laser, we primarily probe devices with pulse lengths ranging from 1 to \SI{1000}{\us} and collection times after the pulse are sufficient to enable decay of the emission (see Methods). The pulses are routed to the sample via a fiber circulator and polarization controller to match the polarization of the cavity mode. We then measure the fluorescence from the sample, back through the lensed fiber and circulator, and directed to either a superconducting nanowire single photon detector (SNSPD), a spectrometer with InGaAs camera, or a photodiode. The one-way coupling efficiency (see Methods) is fairly poor at approximately 15\%, but it is sufficient to probe these particular devices because of the relatively large number of ions in the cavity-coupled ensembles. We also have a fiber wavelength division multiplexer (WDM) that can be added into the path of port 2 of the circulator for off-resonant photoluminescence (PL) measurements using a 1480 nm diode laser. Off-resonant PL measurements on a waveguide-only device (no cavity) reveal a variety of peaks from 1520 nm to 1560 nm, as shown in Fig.~\ref{fig2}b. The largest peak near 1520.5 nm is attributed to substitutional Er in the rutile phase of TiO\textsubscript{2} \cite{Phenicie2019}, whereas the other dominant peak near 1533 nm is attributed to substitutional Er in anatase \cite{Komuro2002}, as confirmed previously with electron diffraction measurements\cite{Singh2022}. This particular sample is polyphase, which has been seen previously for TiO\textsubscript{2} thin films grown at intermediate temperatures~\cite{Singh2022}. The intensity of the substitutional rutile transition---relative to the subsitutional anatase peak---is much stronger in this sample compared to previous thin film results when grown near 520\textdegree C\cite{Singh2022}. The origin of this TiO\textsubscript{2} phase discrepancy is still under investigation. Similarly, the origin of the minor peaks could be due to a variety of factors, such as different localized phases within the TiO\textsubscript{2}, Er$^{3+}$ residing in different substitutional/interstitial sites, or crystal field-split transitions, but additional control samples and experiments are needed.

Resonant photoluminescence excitation (PLE) measurements on the waveguide-only device near 1520.56~nm are performed using a narrowband, tunable CW laser and modulated into pulses with emission from the Er$^{3+}$ detected during a collection window following each pulse. The lifetime decay curve and associated total PLE intensity at that pump wavelength are integrated after numerous pulses (see Methods). Resonant PLE extracts a linewidth of 0.4~nm (Fig.~\ref{fig2}b inset), which is similar to the result from off-resonant pumping. The inhomogeneous linewidth represents the distribution in Er$^{3+}$ transition energies along the entire device that is over \SI{40}{\um} long, and this inhomogeneous linewidth is similar to that measured in predominantly rutile thin film samples using a confocal microscope geometry with a spot size of approximately \SI{1}{\square\um}~\cite{Singh2022}. Given that the inhomogeneous linewidth of the Er$^{3+}$ in rutile transition is relatively broad, it is important to investigate if the homogeneous linewidth of the emitters is sufficiently narrow to enable cavity-based lifetime enhancement (i.e., the ``bad cavity'' limit). To that end, we have performed transient spectral hole burning measurements on a waveguide-only device to find an upper bound on the homogeneous linewidth and spectral diffusion~\cite{Weiss2021} for the rutile transition (Fig.~\ref{fig2}c). Using the pulsed resonant laser PLE scheme described above, we have employed an additional phase modulator ($\phi$EOM, Fig.~\ref{fig2}a) to generate two sidebands each with a frequency detuning, $\Delta$, from the carrier frequency. We can then measure the relative PLE intensity change as $\Delta$ is swept but the laser carrier wavelength is maintained at the rutile transition, 1520.56 nm, and the total pump power is near saturation for the ensemble. If $\Delta$ is larger than the spectral diffusion linewidth of the erbium ions, the ions are no longer pumped to saturation and the total fluorescence increases. If we plot the normalized intensity as a function of laser sideband-carrier detuning (Fig.~\ref{fig2}c), we can fit a inverted Lorentzian lineshape with a half-width half-maximuum (HWHM) of $267\pm17$~MHz, which sets the upper bound on the spectral diffusion-limited homogeneous linewidth\cite{Moerner1988}. While this homogeneous linewidth is substantially narrower than the inhomogeneous linewidth (Fig.~\ref{fig2}b), it is still quite large compared to the natural linewidth of Er$^{3+}$ in rutile TiO\textsubscript{2} (sub kHz). It is possible that further optimization of the top capping layer in these devices can reduce the spectral diffusion of the Er emitters~\cite{Singh2022}.

\begin{figure}
    \includegraphics[width=0.5\textwidth]{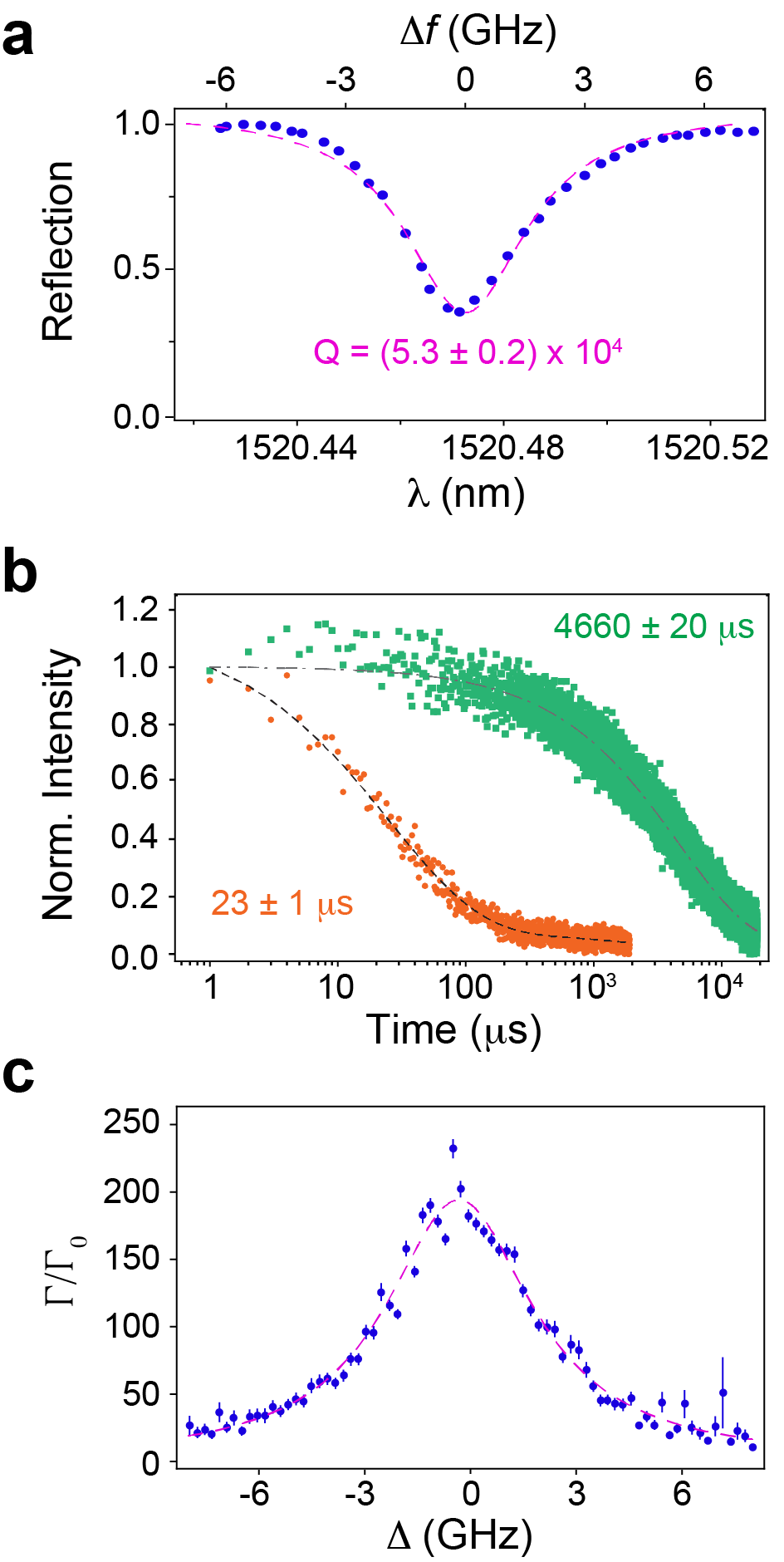}
    \caption{\textbf{Purcell enhancement of Er$^{3+}$:TiO$_2$ ensembles on Si photonic crystal cavities.} (a)~Resonant laser reflection spectrum of a photonic crystal cavity ($\text{T} = \SI{3.1}{\K}$) tuned near the rutile transition, showing a cavity Q near 53,000. The cavity has a corresponding linewidth (full-width half-maximum, FWHM) of \SI{3.71(17)}{\GHz} (top axis). (b) A comparison of the ensemble lifetime of Er$^{3+}$ ions coupled in a waveguide-only device (green circles) versus the cavity-coupled device (orange circles) shown in (a). The corresponding fits give $\tau = \SI{23(1)}{\us}$ (black dashed line) and $\tau = \SI{4660(20)}{\us}$ (gray dashed line), which is a 200-fold reduction in the decay time. (c) A plot of the increase in the ensemble decay rate as a function of the cavity-laser detuning ($\Delta$) when the laser is fixed at \SI{1520.56}{\nm}. This is the same cavity as in (a), which was measured at the start of the detuning experiment. The decay rate enhancement lineshape is also fit to a Lorentzian with a FWHM linewidth of \SI{5.06(17)}{\GHz}.}
    \label{fig3}
\end{figure}

We have the ability to tune the cavity resonance in situ using a combination of controlled gas adsorption and desorption techniques while near base temperature. We use N\textsubscript{2} condensation via a gas nozzle directed at the sample to deposit a thin layer of ice on the cavity which increases the refractive index of the mode, resulting in a redshift of its resonance\cite{Mosor2005}. We can also deterministically desorb the ice through localized heating, whereby we use relatively strong CW laser excitation tuned directly to the cavity resonance to induce two-photon absorption and heating in the Si, but only at the cavity region, which leads to a blueshift of the cavity resonance (see Methods). Furthermore, this heating is sufficiently localized that cavities in neighboring waveguides---separated by only \SI{6}{\um}---can be tuned independently of one another, which can be useful for future two-cavity interference experiments. We can directly measure the spectral position and linewidth of the cavity resonance by detecting the wavelength-dependent reflection of the narrowband laser light using a photodiode. If we perform this measurement close to the rutile transition, we typically see cavity quality factors (Q) near 5$\times 10^{4}$. A prototypical cavity reflection scan at $\text{T} = \SI{3.1}{\K}$ is shown in Fig.~\ref{fig3}a. Through the use of control devices with and without TiO\textsubscript{2} films, the quality factors in general are currently limited by scattering due to the roughness of the top surface of the TiO\textsubscript{2} rather than sidewall roughness of the waveguide or cavity holes from the dry etching, though absorption in spectral regions of high Er$^{3+}$ density also plays a role (see below).

Our resonant pulsed laser measurements reveal the optical lifetime of all ions that couple to the device, which includes ions well-coupled to the cavity, ions poorly coupled to the cavity---whether because of position or polarization---and those that couple only to the bare waveguide. Therefore, we use a stretched exponential function to capture the variety of decay times, where the time constant ($\tau$) in the fit represents the fastest time decay within the ensemble\cite{VandeWalle1996}. Using this technique, the lifetime of the Er ions coupled to a waveguide-only device give a natural lifetime of $\tau = \SI{4660(20)}{\us}$ (Fig. \ref{fig3}b), which is close to the time constant of 5.1 ms measured in implanted/annealed bulk TiO\textsubscript{2} at low temperature\cite{Phenicie2019}. In comparison, the decay time for those ions which couple most strongly to the cavity, when it is tuned onto the rutile transition at 1520.56 nm, show a greatly decreased decay time of $\tau = \SI{23(1)}{\us}$. This represents a decay rate enhancement of 200-fold. 

We next measure the emission rate enhancement as a function of the detuning between the cavity resonance and the Er$^{3+}$ rutile transition. To do so, we perform a systematic redshift of the cavity resonance via gas condensation and measurement of the optical lifetime with the resonant laser fixed at the rutile transition. Figure \ref{fig3}c shows the increase in decay rate relative to the natural decay rate as the cavity resonance is tuned across the transition. The cavity-enhanced emission has a Lorentzian linewidth of \SI{5.06(17)}{\GHz}, which is modestly broader than the cavity linewidth measured via reflection (\SI{3.71(17)}{\GHz}). However, this can be explained by the increase in absorption when the cavity is resonant with the Er$^{3+}$ transition versus the reflection scan which is measured at \SI{90}{\pico\meter} (\SI{11.7}{\GHz}) to the blue of the transition (see Supporting Information). The general agreement of the decay rate enhancement lineshape with that of the cavity suggests that the radiative speedup is due to the cavity rather than non-radiative recombination induced by device fabrication. In addition, the lack of a plateau in Fig.~\ref{fig3}c confirms that the system is in the bad cavity limit, i.e. $\Gamma_{h} < \kappa$, where $\Gamma_{h}$ is the homogeneous linewidth of the emitters and $\kappa$ is the cavity linewidth, in agreement with the transient spectral hole measurement in waveguide-only devices. 

In this Letter, we have shown optical addressing of Er$^{3+}$ ions in rutile phase TiO\textsubscript{2} grown on commercial SOI wafers. We have fabricated top-down 1D photonic crystal cavities from the Si/TiO\textsubscript{2} heterostructure with Purcell enhancements up to 200. Systematic detuning of the cavity resonance with the Er$^{3+}$ transition in rutile phase TiO\textsubscript{2} reveals a good match of the Purcell enhancement to the Lorentzian cavity lineshape indicating the erbium emission rates are limited by the cavity quality factor, not by the emitters' homogeneous linewidths. The next goal for this platform is to isolate and address individual Er ions as single spin quantum memories by cooling down to lower temperatures and applying magnetic fields to address various spin levels. To that end, it is important to push the platform forward in three key areas.

First, it is necessary to reduce the homogeneous and inhomogeneous broadening in this system. The current upper bound of the homogeneous linewidth and spectral diffusion is estimated in the range of a few hundred MHz, and could be attributed to a variety of factors, including trapped charges in the interfacial SiO\textsubscript{x}, fabrication-induced damage, and the presence of mixed phases of TiO\textsubscript{2}. Further improvement of the thin film growth conditions to isolate a single TiO\textsubscript{2} phase and to reduce the interfacial oxide, along with optimization of the buffer layer thicknesses could enable large reductions of the homogeneous and inhomogeneous linewidths, as was seen already in predominantly anatase thin films~\cite{Singh2022}. Second, it is important to drastically reduce the number of Er$^{3+}$ coupled to the cavity in order to spectrally isolate and address single ions. By further reducing the metallic Er source flux and restricting doping to a single delta-doped TiO\textsubscript{2} layer, intentional doping of these films to the single ion level is straightforward via molecular beam deposition. Additional improvements to the homogeneous linewidth may also be achieved naturally with the reduction in the Er doping density~\cite{Singh2022} that would be needed to address single ions. Importantly, telecom quantum technologies based on this platform will be enabled by improving the photonic device performance. Most notably, the current fiber-waveguide coupling efficiency will need to be increased through improved mode-matching via cladding layers\cite{Pu2010} or a full undercut of the Si inverse tapered waveguides~\cite{Meenehan2014}. A reduction in the TiO\textsubscript{2} surface roughness can lead to improvements in cavity Q for higher photon rates. The aforementioned reduction in Er doping density will also decrease parasitic absorption that is affecting cavity quality factors at low temperature in current devices. 

This demonstration represents an exciting first step towards a scalable on-chip quantum light source and memory device platform grown on an SOI wafer. This telecom-ready device architecture can be integrated with standard Si photonics and MEMS foundry process lines, enabling integration with other needed photonic elements such as on-chip filters, phase shifters, and beamsplitters. Furthermore, the Si platform is amenable to combination with other important cryogenic quantum technologies such as on-chip SNSPDs, superconducting qubits, semiconductor quantum dot qubits, and microwave-to-optical transducers in order to enable powerful quantum computational and communication nodes. With further improvements to the materials and device fabrication, the scalable nature of this Er-doped TiO\textsubscript{2}-on-SOI platform can meet the future demands for large numbers of communication qubits and quantum light sources needed for fault-tolerant long-distance quantum networking.

\section*{Methods}
\textbf{Film growth, device fabrication, and electron microscopy}\\
TiO\textsubscript{2} thin films are grown on diced pieces from 8" commercial SOI wafers (SOITEC). The Si device layer is lightly boron-doped with a resistivity of \SI{10}{\ohm\cdot\cm}. The TiO\textsubscript{2} thin film growth conditions and methods are outlined in Singh et al.~\cite{Singh2022}. For this particular film the metallic Er source temperature is at 900\textdegree~C, with an expected doping density of 35 ppm of Er$^{3+}$, to be later confirmed with secondary ion mass spectrometry. Following growth, all device fabrication is performed in the Center for Nanoscale Materials cleanroom at Argonne National Laboratory. Device patterning is performed using electron beam lithography (JEOL 8100). The etch mask is a combination of electron-beam resist (ZEP 520A) and a plasma enhanced chemical vapor deposition (PECVD, Oxford PlasmaLab 100) SiO\textsubscript{2} hardmask. Fluorine-based etching is used for mask transfer to the SiO\textsubscript{2}, chlorine-based etching is used to etch through the TiO\textsubscript{2} layers, and HBr/O\textsubscript{2} is used to etch through the Si device layer. All etching is performed in an Oxford PlasmaLab 100 inductively coupled plasma (ICP) reactive ion etcher (RIE). Waveguide cleaving along alignment marks is performed under an optical microscope (LatticeAx 420, LatticeGear). A full device fabrication flow is provided in the Supporting Information. For TEM cross-section analysis, the opposite side of the cleaved chip shown in~Fig.~\ref{fig1}d-e was mechanically ground to a thickness of \SI{20}{\um}, followed by Ar ion milling to electron transparency. High-resolution TEM was performed at 200 kV using the Argonne chromatic aberration-corrected TEM (ACAT) at the Center for Nanoscale Materials.\\

\noindent\textbf{Optical Characterization}\\
Measurements are performed in a Montana Instruments S100 closed-cycle cryostat. The AR coated SMF-28 lensed fiber (TSMJ-X-1550-9/125-0.25-18-2.5-14-3-AR, OZ Optics) is mounted on an Attocube nanopositioner with 5~mm of x-y-z travel. The base temperature of the cold finger directly underneath the sample is measured to be \SI{3.1}{\K}. Resonant experiments are performed with a tunable telecom laser (CTL 1550, Toptica). The wavemeter used for wavelength measurement is a High Finesse WS8-10 with SLR-1532 calibration laser. The one-way coupling efficiency is estimated by measuring the ratio of the reflected laser power from a cavity device slightly off resonance to the laser power into the lensed fiber, and then taking the square root of this ratio. Our efficiency estimate of 15\% encompasses all losses from the lensed fiber vacuum feedthrough to the device: including fiber bend losses, scattering along the inverse taper on the waveguide, and the fiber-waveguide coupling itself. However, due to the limitation that the waveguide be sufficiently wide to enable guiding and not lose light into the buried oxide layer, the mode mismatch between the waveguide and optical spot (\SI{\sim 2.5}{\um}) of the lensed fiber is currently the dominant source of loss.

Pulses used for resonant laser pumping are generated with fiber acousto-optic modulators, AOMs (FiberQ, Gooch and Housego). The collective on/off ratio of these three modulators is greater than 150 dB. The typical pulses for waveguide-only devices are 0.1-\SI{1}{\ms} long (CW laser power of 0.3-\SI{3}{\uW} at the device) with collection times of 10-\SI{40}{\ms}. Typical pulses for cavity-coupled ions are 1-\SI{10}{\us} long (CW laser power of \SI{\sim 50}{\nano W} at the device) with collection times of 1-\SI{4}{\ms}. The phase electro-optic modulator (MPZ-LN-10, iXblue) is driven with a vector signal generator (SG396, SRS) for sideband generation at a specified detuning from the laser carrier frequency. Single photon detection was performed with a SNSPD from Quantum Opus inside a second cryostat, which is designed for operation near \SI{1550}{nm} with a dark/background count rate of 50~Hz and external quantum efficiency of \SI{83}{\percent}. In the collection path, an additional fiber AOM (MT80-IIR30-Fio-PM0.5-J1-A-Ic2, AA Opto Electronic) is used to protect the SNSPD from direct laser exposure to mitigate transients in the detector signal. Single photon counting is performed using a dedicated time tagger (quTAG, qutools). Two voltage controlled attenuators (V1550PA, Thorlabs) are used in series to control the incident laser power for resonant PLE experiments as well as in situ resonant laser cavity tuning. It is important to note that the fiber circulator is particularly useful---instead of a 99:1 fiber beam splitter, for example---because it allows for sufficiently high input powers at the device (\SI{\sim 10}{\uW}) after accounting for modulator and fiber-to-device losses. This enables resonant laser tuning (2-photon absorption in Si) of the cavity without an additional amplifier. 

Off-resonant PL measurements were performed by inserting a 1480 nm diode pump laser (QFBGLD-1480-300, QPhotonics) via a 1480/1550 nm fiber wavelength division multiplexer (WD1450A, Thorlabs). Erbium PL was detected after long-pass filtering (FELH1500, Thorlabs) using a spectrometer (IsoPlane SCT320, Princeton Instruments) equipped with a Pylon-IR liquid-nitrogen cooled InGaAs camera. The spectrometer has a 600 grooves/mm grating, and the resolution of \SI{\sim 0.1}{\nm} was confirmed by measuring a narrow linewidth laser.

\section*{Supporting Information} 
A detailed device nanofabrication flow and additional experimental cavity reflection measurements on parasitic absorption near the Er$^{3+}$ transition in rutile TiO\textsubscript{2} are provided in the Supporting Information.

\section*{Acknowledgements}
The authors would like to thank D. Czaplewski, C.S. Miller, and R. Divan for assistance with fabrication. They would also like to thank G. Wolfowicz and M. Raha for constructive feedback. This work, including materials growth and optical characterization, was primarily supported by the Q-NEXT Quantum Center, a U.S. Department of Energy, Office of Science, National Quantum Information Science Research Center, under Award Number DE-FOA-0002253. Nanofabrication and electron microscopy work performed at the Center for Nanoscale Materials, a U.S. Department of Energy Office of Science User Facility, was supported by the U.S. DOE, Office of Basic Energy Sciences, under Contract No. DE-AC02-06CH11357. Additional support for cryogenic and optical infrastructure development was provided by U.S. Department of Energy, Office of Science; Basic Energy Sciences; Materials Sciences and Engineering Division. Additional support for growth capabilities was provided by the Center for Novel Pathways to Quantum Coherence in Materials, an Energy Frontier Research Center funded by the U.S. Department of Energy, Office of Science, Basic Energy Sciences under Award No. DE-AC02-05CH11231.

\clearpage
\bibliography{TiO2_devices}

\end{document}